\documentclass{PoS}
\usepackage{graphicx}
\usepackage{url}

\newcommand\apjl{Astrophysical Journal, Letters}

\newcommand\apjs{Astrophysical Journal, Supplement}

\newcommand\aap{Astronomy and Astrophysics}

\title{Spectral analysis of the blazars Markarian 421 and Markarian 501 with the HAWC Gamma-Ray Observatory}

\ShortTitle{Spectral analysis of Markarian 421 and Markarian 501 with HAWC}

\author{Sara Couti\~no de Le\'on, \speaker{Alberto Carrami\~nana Alonso}, Daniel Rosa-Gonz\'alez and Anna Lia Longinotti for the HAWC collaboration\footnote{For collaboration list see PoS(ICRC2019)1177 and https://www.hawc-observatory.org/collaboration/icrc2019.php}\\

Instituto Nacional de Astrof\'isica, \'Optica y Electr\'onica, Puebla, Mexico.\\
        E-mail: \email{sara@inaoep.mx}, \email{alberto@inaoep.mx}, \email{danrosa@inaoep.mx}, \email{annalia@inaoep.mx}}

\abstract{The High Altitude Water Cherenkov (HAWC) Gamma-Ray Observatory surveys the very high energy sky in the $\sim300$ to 100 TeV energy range and has detected two high-synchrotron-peaked BL Lacertae objects, Markarian 421 and Markarian 501 in a period of time of 837 days between June 2015 and December 2017. In this work we present the detailed time-average spectral analysis. Using an extragalactic background light model, we address the difference in the intrinsic spectral properties between the two blazars above 1 TeV, preliminary results show that the intrinsic spectrum of Mrk 421 is better described by a power-law with an exponential energy cut-off function with photon index $\alpha_{421}=2.22\pm0.10$ and energy cut-off $Ec_{421}=5.24\pm1.02$ TeV, and for Mrk 501 the intrinsic spectrum is well described by a power law with spectral index $\alpha_{501}=2.40\pm0.06$, without requiring an energy cut-off.}

\FullConference{36th International Cosmic Ray Conference -ICRC2019-\\
		July 24th - August 1st, 2019\\
		Madison, WI, U.S.A.}

\begin{document}

\section{Introduction}
Most extragalactic sources of very high-energy (VHE) radiation are associated to active galactic nuclei (AGN), characterized by ultra-relativistic jets escaping from a super massive black hole (SMBH) in the central region of the galaxies. Two of the most studied and brightest sources in the extragalactic TeV sky are Markarian 421 (Mrk 421) and Markarian 501 (Mrk 501) with a redshift of $z=0.031$ and $z=0.034$, respectively. They were first detected at VHE by the Whipple Observatory, the first one in 1992 \cite{Mrk421discover} and the second one in 1996 \cite{Mrk501discovery}. They were classified as high-synchroton-peaked (HBL) BL Lac objects \cite{Mrk421HSP} and their gamma-ray spectrum has been measured with many imaging atmospheric Cherenkov telescopes (IACTs). They are continuously monitored by the \textit{Fermi}-LAT whose latest report is in \textit{Fermi} Large Area Telescope Third Source Catalog \cite{3fgl}.

The spectral energy distribution of HBL has two components, one at low energies is originated via synchrotron radiation of relativistic electrons, whereas the high-energy that is commonly attributed to inverse Compton scattering (IC); however the nature of the high energy component  is still under debate \cite{Jones, Mannheima, Mannheimb}. For this reason, Mrk 421 and Mrk 501 are often studied to constrain emission blazar models by separating their intrinsic properties and the attenuation effects due to the extragalactic background light (EBL). In this work we present the preliminary results of the VHE observations of both sources as a result of  837 days of observations with the High Altitude Water Cherenkov (HAWC) Gamma-Ray Observatory in order to address the difference in their intrinsic spectral properties.

\section{Data}\label{Data}

The HAWC Observatory is a ground-based TeV gamma-ray detector in the state of Puebla, Mexico at an altitude of 4100 m a.s.l. The detector continuously measures the arrival time and direction of cosmic and gamma-ray primaries within its 2 sr field of view. It is most sensitive to gamma-ray energies ranging from 300 GeV - 100 TeV. Cuts on the data can be applied to differentiate gamma-ray air showers from the large cosmic-ray background. A reconstruction of the air shower in the detector is performed to obtain the event's size and the direction of the primary $\gamma$-ray. For more details about the detector performance read \cite{Crab,catalog}.

The data used for this analysis goes from June 2015 to December 2017 comprising 837 days of effective exposure. The data is divided into bins which definition depends on the fraction of photomultiplier tubes that are triggered in each event; however this method does not take into account energy assignment variables so the energy of the gamma-ray events is estimated through a ground parameter method that measures the charge 40 meters from the air shower axis of each event and the estimated energy is found performing a fit using an empirical function that matches the simulated events, this way the bins are sub-divided into quarter-decade energy bins in the 0.316-100 TeV range. Due to the poor energy resolution below 1 TeV, we performed the fits above this energy. For more details about the energy estimator performance see \cite{crab-gp}.

\section{Analysis}\label{Analysis}

A forward-folding method is performed to fit the spectral shape of the sources using the HAWC maximum-likelihood framework (LiFF) described in \cite{LiFF}. Given a source spectral model, the data and background maps are convolved with the detector response and LiFF computes a binned Poisson log-likelihood value. The test statistic ($TS$) is computed using the likelihood of the observations, estimated using a background-only model (null hypothesis, $H_0$), and the signal-plus-background model (alternative hypothesis, $H_1$). $TS$ is defined as 
\begin{equation} \label{TS}
TS = 2 \ln\frac{\mathcal{L}(H_1)}{\mathcal{L}(H_0)},
\end{equation}
where $\mathcal{L}$ is the likelihood function. To determine a source spectrum, the $TS$ is numerically maximized by iteratively changing the input parameters, yielding those values that have the highest likelihood of describing the observed data for the point source model assumption. Using Wilks' Theorem \cite{wilkstheorem} the significance is $\sigma=\pm \sqrt{TS}$ and according to \cite{catalog}, we consider a source detected when $\sigma>5$.

In order to estimated an intrinsic spectrum, the input spectral model is assumed to be the intrinsic one which is then attenuated using an EBL model. This attenuated spectral model is the one convolved with the detector response to be then compared with the observed counts. This way, the output parameters correspond to the intrinsic ones. The EBL model used to perform the fits in this work is from \cite{Gil}. 

For this work, we tested two spectral shapes, a single power law (PL, equation \ref{PL}) and a power law with an exponential energy cut-off (PL+CO, equation \ref{PL+CO}). 

\begin{equation}\label{PL}
\frac{dN}{dE} = N_0 \left(\frac{E}{E_0}\right)^{-\alpha}\times \exp(-\tau),
\end{equation}

\begin{equation}\label{PL+CO}
\frac{dN}{dE} = N_0 \left(\frac{E}{E_0}\right)^{-\alpha}\times\exp\left(\frac{-E}{E_c}\right)\times \exp(-\tau),
\end{equation}
where $N_0$ is the flux normalization $[\mbox{TeV}^{-1}\mbox{cm}^{-2}\mbox{s}^{-1}]$, $E_0$ is the pivot energy fixed at 1 TeV, $\alpha$ is the spectral index, $E_c$ is the energy cut-off $[\mbox{TeV}]$, and $\tau$ is the opacity value given by EBL models, and which is an increasing function of $E$ and the source redshift, $z$. 

Depending on the $TS$ values in the global fit using all the available energy bins, a preferred spectral shape is chosen. As described in \cite{crab-gp}, the process to estimate flux points is performing individual fits in each energy bin where only $N_0$ is free to vary and $\alpha$ and $E_c$ are fixed using the resulting values from the global fit. If a fit from an individual energy bin has a $TS<25$, an upper limit at a $95\%$ confidence interval is set following \cite{ULFelman}.

\section{Results}\label{Results}

\begin{figure}
\centering
\includegraphics[scale=0.6]{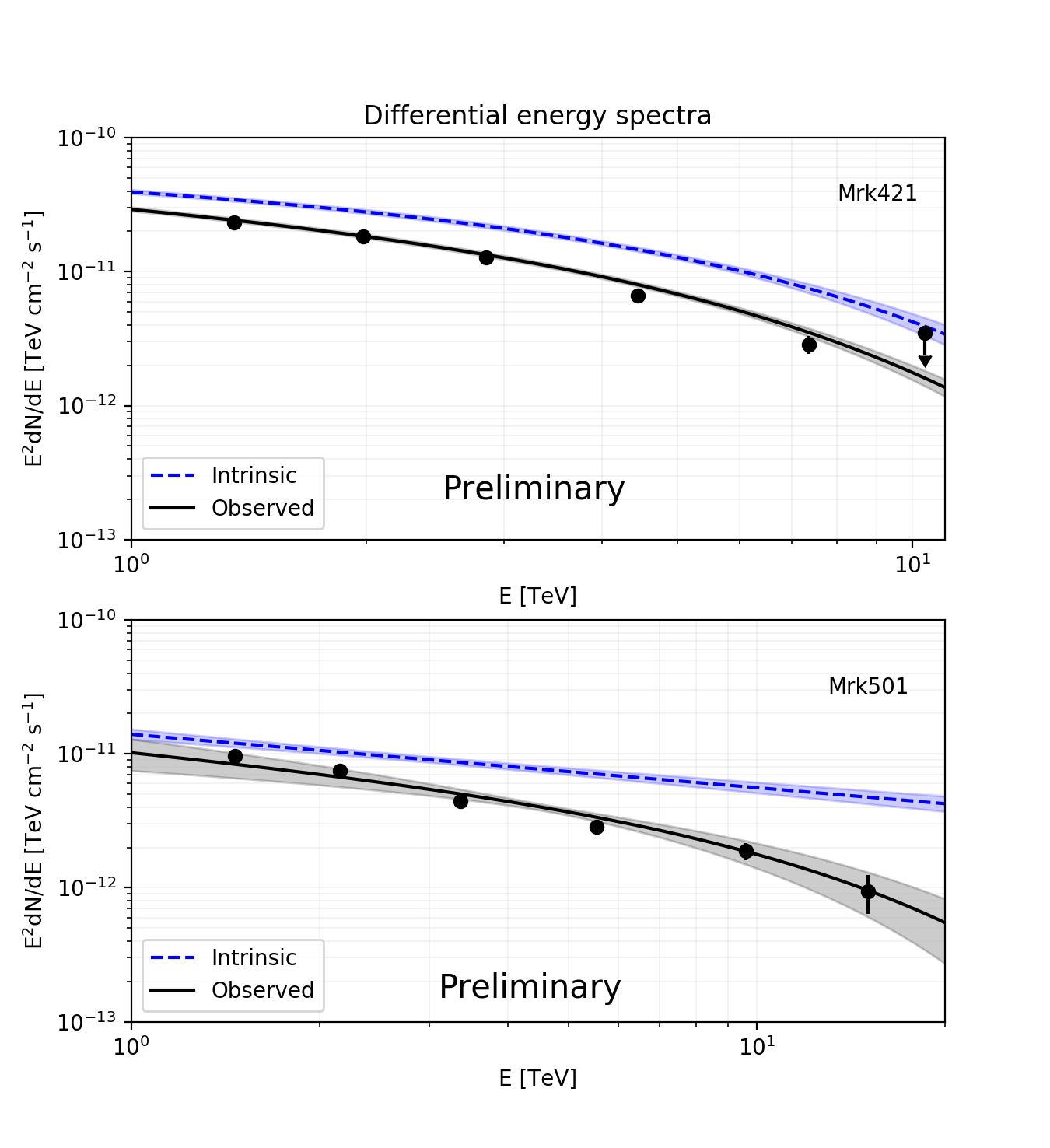}
\caption{Differential energy spectra of Mrk 421 (top) and Mrk 501 (bottom). The observed (black line and dots) spectra is the best fit to the data and the intrinsic spectra (blue dashed line) is obtained using the EBL model from \cite{Gil}. The black points are the result of individual fits in each energy bin as described in section \ref{Analysis}.}\label{spectra}
\end{figure}

The model that best describes the intrinsic spectrum of the sources is a PL+CO for Mrk 421; however for Mrk 501 the spectral fit has very similar values of $TS$ for both a PL and a PL+CO spectral models but the fitted energy cut-off is larger than 700 TeV, so that within the HAWC energy range the intrinsic spectrum can be modeled with a single PL. The resulting spectral parameters are given in Table \ref{parameters} where the corresponding uncertainties are statistical only. Figure \ref{spectra}  show the best fits for the observed (black) and intrinsic (blue) spectra of both sources.

\begingroup
\renewcommand{\arraystretch}{1.3} 
\begin{table}
\centering
\caption{Fitted spectral parameters for Mrk 421 and Mrk 501 following the method described in section \ref{Analysis}.}\label{parameters}
\begin{tabular}{c c c c c}
\hline \hline
   & $\sqrt{TS}$ &$N_0$ & $\alpha$ & $E_c$\\
   &  & \small{$[\mbox{TeV}^{-1}\mbox{cm}^{-2}\mbox{s}^{-1}]$} & & \small{[TeV]}\\
\hline
Mrk 421    & 48 & $(4.77 \pm 0.25) \times10^{-11}$ & $2.22 \pm 0.10$  & $5.24 \pm 1.02$  \\
Mrk 501 	& 22 & $(1.40 \pm 0.12) \times10^{-11}$ & $2.40 \pm 0.06$  & $\infty$  \\
\hline \hline
\end{tabular}
\end{table}
\endgroup

\section{Summary}\label{Conclusions}

We present the VHE spectra of Mrk 421 and Mrk 501, the nearest BL Lacs, with HAWC. The data set comprises 837 transits and the spectral analysis was performed using an energy estimator based on the measurement of the charge at the ground to divide the data into energy bins \cite{crab-gp}, and the HAWC maximum-likelihood framework \cite{LiFF}. 

The differential energy spectra of each source is also EBL-corrected using the model from \cite{Gil}, and the preliminary results show that Mrk 421 has a curved shape which may indicate photon-photon attenuation inside the source, while Mrk 501 is well described by a simple power law. These differences prompt a deeper exploration of the intrinsic nature of both sources through blazar emission models in order to determine the physical processes and the nature of the particles that produce radiation at very high energies.

\bibliographystyle{JHEP}
\bibliography{biblio}

\providecommand{\href}[2]{#2}\begingroup\raggedright\begin{thebibliography}{10}

\bibitem{Mrk421discover}
M.~{Punch}, C.~W. {Akerlof}, M.~F. {Cawley}, M.~{Chantell}, D.~J. {Fegan},
  S.~{Fennell} et~al., \emph{{Detection of TeV photons from the active galaxy
  Markarian 421}}, \href{https://doi.org/10.1038/358477a0}{\emph{\nat}
  {\bfseries 358} (Aug., 1992) 477}.

\bibitem{Mrk501discovery}
J.~{Quinn}, C.~W. {Akerlof}, S.~{Biller}, J.~{Buckley}, D.~A. {Carter-Lewis},
  M.~F. {Cawley} et~al., \emph{{Detection of Gamma Rays with E $>$ 300 GeV from
  Markarian 501}}, \href{https://doi.org/10.1086/309878}{\emph{\apjl}
  {\bfseries 456} (Jan., 1996) L83}.

\bibitem{Mrk421HSP}
A.~A. {Abdo}, M.~{Ackermann}, I.~{Agudo}, M.~{Ajello}, H.~D. {Aller}, M.~F.
  {Aller} et~al., \emph{{The Spectral Energy Distribution of Fermi Bright
  Blazars}}, \href{https://doi.org/10.1088/0004-637X/716/1/30}{\emph{\apj}
  {\bfseries 716} (June, 2010) 30--70},
  [\href{https://arxiv.org/abs/0912.2040}{{\ttfamily 0912.2040}}].

\bibitem{3fgl}
F.~{Acero}, M.~{Ackermann}, M.~{Ajello}, A.~{Albert}, W.~B. {Atwood},
  M.~{Axelsson} et~al., \emph{{Fermi Large Area Telescope Third Source
  Catalog}}, \href{https://doi.org/10.1088/0067-0049/218/2/23}{\emph{\apjs}
  {\bfseries 218} (June, 2015) 23},
  [\href{https://arxiv.org/abs/1501.02003}{{\ttfamily 1501.02003}}].

\bibitem{Jones}
T.~W. {Jones}, S.~L. {O'dell} and W.~A. {Stein}, \emph{{Physics of Compact
  Nonthermal Sources. I. Theory of Radiation Processes}},
  \href{https://doi.org/10.1086/152724}{\emph{\apj} {\bfseries 188} (Mar.,
  1974) 353--368}.

\bibitem{Mannheima}
K.~{Mannheim}, \emph{{{$\gamma$} rays and neutrinos from a powerful cosmic
  accelerator}}, \href{https://doi.org/10.1103/PhysRevD.48.2408}{\emph{\prd}
  {\bfseries 48} (Sept., 1993) 2408--2414},
  [\href{https://arxiv.org/abs/astro-ph/9306005}{{\ttfamily
  astro-ph/9306005}}].

\bibitem{Mannheimb}
K.~{Mannheim}, \emph{{The proton blazar}}, {\emph{\aap} {\bfseries 269} (Mar.,
  1993) 67--76}, [\href{https://arxiv.org/abs/astro-ph/9302006}{{\ttfamily
  astro-ph/9302006}}].

\bibitem{Crab}
A.~U. {Abeysekara}, A.~{Albert}, R.~{Alfaro}, C.~{Alvarez}, J.~D.
  {{\'A}lvarez}, R.~{Arceo} et~al., \emph{{Observation of the Crab Nebula with
  the HAWC Gamma-Ray Observatory}},
  \href{https://doi.org/10.3847/1538-4357/aa7555}{\emph{ApJ} {\bfseries 843}
  (July, 2017) 39}, [\href{https://arxiv.org/abs/1701.01778}{{\ttfamily
  1701.01778}}].

\bibitem{catalog}
A.~U. {Abeysekara}, A.~{Albert}, R.~{Alfaro}, C.~{Alvarez}, J.~D.
  {{\'A}lvarez}, R.~{Arceo} et~al., \emph{{The 2HWC HAWC Observatory Gamma-Ray
  Catalog}}, \href{https://doi.org/10.3847/1538-4357/aa7556}{\emph{ApJ}
  {\bfseries 843} (July, 2017) 40},
  [\href{https://arxiv.org/abs/1702.02992}{{\ttfamily 1702.02992}}].

\bibitem{crab-gp}
{HAWC Collaboration}, A.~U. {Abeysekara}, A.~{Albert}, R.~{Alfaro},
  C.~{Alvarez}, J.~D. {{\'A}lvarez} et~al., \emph{{Measurement of the Crab
  Nebula at the Highest Energies with HAWC}}, {\emph{arXiv e-prints} (May,
  2019) arXiv:1905.12518}, [\href{https://arxiv.org/abs/1905.12518}{{\ttfamily
  1905.12518}}].

\bibitem{LiFF}
P.~W. {Younk}, R.~J. {Lauer}, G.~{Vianello}, J.~P. {Harding}, H.~A. {Ayala
  Solares}, H.~{Zhou} et~al., \emph{{A high-level analysis framework for
  HAWC}},  in \emph{34th International Cosmic Ray Conference (ICRC2015)},
  vol.~34 of \emph{International Cosmic Ray Conference}, p.~948, July, 2015.

\bibitem{wilkstheorem}
S.~S. {Wilks}, \emph{{The Large-Sample Distribution of the Likelihood Ratio for
  Testing Composite Hypotheses}}, {\emph{{Ann. Math. Statist.}} {\bfseries 9}
  (1938) 60--62}.

\bibitem{Gil}
R.~C. {Gilmore}, R.~S. {Somerville}, J.~R. {Primack} and A.~{Dom{\'{\i}}nguez},
  \emph{{Semi-analytic modelling of the extragalactic background light and
  consequences for extragalactic gamma-ray spectra}},
  \href{https://doi.org/10.1111/j.1365-2966.2012.20841.x}{\emph{\mnras}
  {\bfseries 422} (June, 2012) 3189--3207},
  [\href{https://arxiv.org/abs/1104.0671}{{\ttfamily 1104.0671}}].

\bibitem{ULFelman}
G.~Feldman and R.~Cousins, \emph{Unified approach to the classical statistical
  analysis of small signals}, {\emph{Phys. Rev. D} {\bfseries 57(7)} (1998)
  3873--3889}.

\end{thebibliography}\endgroup

\end{document}